\documentclass[12pt]{iopart}
\usepackage{iopams}  

\usepackage{graphicx}  
\usepackage{dcolumn}   
\usepackage{bm}        
\usepackage{xspace}
\usepackage{xcolor}
\usepackage{float}
\usepackage{subcaption}

\usepackage [T1]{fontenc}
\usepackage [utf8]{inputenc}
\usepackage[paperwidth=210mm,paperheight=297mm,centering,hmargin=2cm,vmargin=2.5cm]{geometry}
\usepackage{multirow}

\usepackage[square,numbers,sort&compress]{natbib}
\usepackage[colorlinks]{hyperref}
\newcommand{\text}{\mathrm}
\newcommand{\eqref}[1]{(\ref{#1})}

\graphicspath{{./figures/}{./}}

\begin{document}

\title{Multiscaling in the $3D$ critical site-diluted Ising ferromagnet}

\author{E.~Marinari$^{1}$,V.~Martin-Mayor$^{2}$, G. Parisi$^{1}$, F. Ricci-Tersenghi$^{1}$, J.J.~Ruiz-Lorenzo$^{3}$}

\address{$^1$ Dipartimento di Fisica, Sapienza  Universit\`a di Roma and CNR-Nanotec,
  Rome unit, and INFN, Sezione di Roma, I-00185 Rome,  Italy}
\address{$^{2}$ Departamento de F\'\i{}sica
  Te\'orica, Universidad Complutense, 28040 Madrid,
  Spain}
\address{$^{3}$ Departamento de F\'{\i}sica and Instituto de Computaci\'on   Cient\'{\i}fica Avanzada (ICCAEx), Universidad de Extremadura, 06006 Badajoz,
  Spain}

  \begin{abstract}
\noindent 
We have studied numerically the appearance of multiscaling behavior in the three-dimensional ferromagnetic Ising site diluted model, in the form of a multifractal distribution of the decay exponents for the spatial correlation functions at the critical temperature.  We have computed the exponents of the long-distance decay of higher moments of the correlation function, up to the 10th power, by studying three different quantities: global susceptibilities, local susceptibilities and correlation functions. We have found very clear evidences for multiscaling behavior.
\end{abstract}

\date{\today}
\maketitle

\newpage
\tableofcontents
\newpage

\section{Introduction}

The extension of Statistical Mechanics to the study of systems with quenched disorder is of paramount importance. Many efforts have been devoted to this task in the last decades, and many important results have been established. Universality is here the main key. Critical exponents can take, for pure systems, few values, that only depend on crucial features of the system, like its dimensionality, the nature of its fields and the symmetries of the Hamiltonian. How does this idea extends to system where the Hamiltonian is characterized by quenched disorder has been discussed at length and is still, at least partially, an open problem.

The Diluted ferromagnetic Ising Model (DIM in the following) is one of the crucial prototypes on which the entrance of quenched disorder in Statistical Mechanics is analyzed. Here even an infinitesimal amount of dilution (that can be a site or a link dilution, leading to models showing basically the same physical behavior) on a pure model, with a positive specific heat critical exponent, will change universality class and the critical exponents~\cite{harris:74,cardy:96}. 

Here in order to further clarify crucial properties of the DIM we will be interested in the behavior of correlation functions and of susceptibilities. In a second order phase transition a correlation length $\xi$ diverges at the critical point $T_c$ when the volume $V$ of the system diverges. When the position of the critical point is known we can analyze correlation functions $C$ there, where they are expected to scale as a power law of the distance, with an exponent, say, $\tau$. An usual scaling law would suggest that the $q$-th moment of a correlation function $C(r)$ would scale with an exponent $\tau(q)$ equal to $ q \tau(1) $ (see later for details and precise definitions). 

We will find here that in a three dimensional, $D=3$ Diluted Ising Model, this is not true, and we get what is called a \textit{multi-fractal} behavior~\cite{harte:01,barabasi:09}. We will determine with high accuracy the values of these exponents, and show numerically that they do not obey the relation $\tau(q)=q \tau(1)$.

This multiscaling behavior of the correlation functions has been thoroughly studied in the context of two-dimensional ferromagnetic spin models with quenched disorder (Potts models with more than two states and  site or link dilution) using conformal field techniques and numerical simulations~\cite{ludwig:89,dotsenko:98,lewis:98,palagny:00,chatelain:01,berche:02,monthus:09}. In another context, we can also quote the multiscaling behavior which appears in the strong amplitude fluctuations of the wave functions in the Anderson localization~\cite{evers:08}.

To find such a results in a $3D$ "simple" disordered system (simple as opposed to spin glasses, where the same fact has been recently established numerically for all values of the temperature in the broken phase~\cite{janus:23c}) is in some sense unexpected, but indeed some important theoretical results by Davis and Cardy had clarified, already many years ago, the plausibility of this fact~\cite{cardy:99,davis:00}. Conformal invariance allows crucial developments about critical scaling, starting from two dimensional systems~\cite{ludwig:89}. By applying to this context renormalization group and an $\epsilon$ expansion Davis and Cardy were able to show that in the DIM one expects multifractality in $2+\epsilon$ dimensions~\cite{davis:00}. 

Our numerical findings, that we support with theoretical arguments, are consistent with Davis and Cardy's computation~\cite{davis:00} (see also  \ref{app:daviscardy}). As many results in the field explain, there are indeed many factors that have to be considered: logarithmic corrections also appear in these kinds of contexts, and we have been able to use numerics to clearly show that in this case we are observing a real, bona fide multifractality. We will discuss this in detail in the following.

The structure of the paper is the following.
We shall first define our model (Sect.~\ref{sect:model}), describe the numerical approach we follow (Sect.~\ref{sect:simulation}), define the observable quantities that we compute (Sect~\ref{sect:observables}) and then explain how we compute these quantities (Sect~\ref{sect:observables-computation}). Next, we shall show our numerical results in Sect.~\ref{sect:results}. Finally, we shall discuss our results and draw some conclusions in Sect.~\ref{sect:conclusions}. The paper is complemented with an Appendix, where we remind the reader of relevant theoretical results.

\section{Model, simulations, and observables}

\subsection{The model}\label{sect:model}
We have considered the $D=3$ diluted Ising model defined on a cubic lattice with periodic boundary conditions, linear size $L$ and volume $V=L^D$. The Hamiltonian of the model, in zero magnetic field, has the form
\begin{equation}\label{eq:H-def}
   {\cal H}=-\sum_{\langle\boldsymbol{x},\boldsymbol{y}\rangle} \epsilon_{\boldsymbol{x}} 
   \epsilon_{\boldsymbol{y}} s_{\boldsymbol{x}} s_{\boldsymbol{y}}\,, 
\end{equation}
where $s_{\boldsymbol{x}}$ are Ising variables and $\epsilon_{\boldsymbol{x}}$  are statistically independent quenched random variables, which take the value 1 with probability $p$ and the value 0 with probability $1-p$. A set of $\{\epsilon_{\boldsymbol{x}}\}$ defines a \emph{sample}.
The sum in Eq.~\eqref{eq:H-def} runs over all pairs of lattice nearest-neighbors. As usual, we denote with $\langle (\cdots)\rangle$ the thermal average for the $\{s_{\boldsymbol{x}}\}$, that is computed for a \emph{fixed} set of disorder variables $\{\epsilon_{\boldsymbol{x}}\}$.
The average over the samples, $\overline{(\cdots)}$, is only taken after thermal mean values have been computed for each sample.

\subsection{Our numerical simulations}\label{sect:simulation}

We have investigated the model defined in Eq.~\eqref{eq:H-def} through equilibrium numerical simulations. Specifically, we have brought to thermal equilibrium our samples by combining Wolff' single-cluster algorithm~\cite{wolff:89}, with a local Metropolis method. The remarkable effectiveness of this combination of simulations algorithms in DIM simulations was demonstrated long ago~\cite{ballesteros:98, ballesteros:98b}. Specifically, our elementary Monte Carlo steps consisted on $L$ single-cluster updates, followed by a sequential full-lattice Metropolis update.

In particular, we have fixed the (average) spin density to $p=0.8$, because for this value the scaling corrections are very small, even negligible given our statistical errors: the model is almost governed by a  \textit{perfect action}~\cite{ballesteros:98b, hasenbusch:07}. Furthermore, we have performed all our simulations at  the infinite volume critical inverse temperature~\cite{hasenbusch:07}: 
$\beta_c^{(p=0.8)}=0.2857429(4)$. For future use, we note that the anomalous dimension of the field is $\eta=0.036(1)$~\cite{hasenbusch:07}. Further details about our simulations can be found in Table~\ref{tab:stat}.

\subsection{Definition of main observables} \label{sect:observables}
In the next three paragraphs we describe the different observables used to analyzed the multiscaling properties of our model~\eqref{eq:H-def}.

\subsubsection{Correlations}
The relevant correlation functions are here
\begin{equation}
    C_q(r)=\frac{1}{p V} \sum_{\boldsymbol{x}}   \overline{ \langle  \epsilon_{\boldsymbol{x}}  s_{\boldsymbol{x}}
     \epsilon_{\boldsymbol{x+r}} s_{\boldsymbol{x+r}}\rangle^q}  \,. 
\label{eq:defCK}
\end{equation}
As usual, we are assuming that rotational invariance is recovered in the scaling limit $\xi\gg 1$, and we indicate only the dependence of $C_q(r)$ on the length $r$ of the displacement vector $\boldsymbol{r}$.

In order to study  multiscaling behavior, we introduce  the $\zeta(q)$ and $\tau(q)$ exponents  through the following relations:
\begin{equation}
 C_q(r) \sim \frac{1}{r^{\tau(q)}} \sim [C_1(r)]^{\zeta(q)}\,.   \label{eq:defTheta}
\end{equation}
It follows that
$\tau(q)$ and $\zeta(q)$ are connected by the relation
\begin{equation}
    \tau(q)=(D-2+\eta) \zeta(q)\,.
    \label{eq:tautheta}
\end{equation}
because
\begin{equation}
C_1(r) \sim \frac{1}{r^{D-2+\eta}}\,.
\end{equation}
By definition, $\zeta(1)=1$ and  $\tau(1)=D-2+\eta$. In the absence of multiscaling behavior one would have $\tau(q)= (D-2+\eta) q$ or, equivalently,  $\zeta(q)=q$ (see  \ref{app:scaling} for more details). Instead, we shall find that $\zeta(q)<q$ whenever $q>1$.

Finally, in order to compute the $\zeta$ exponent, minimizing the scaling corrections,  we will analyze the ratio
\begin{equation}
\label{eq:beta}
 \frac{C_q}{C_{1}^q} \sim [C_{1}]^{\zeta(q)-q}\,.   
\end{equation}

\subsubsection{Global susceptibilities}
We define the $\chi_q$ susceptibilities as
\begin{equation}
    \chi_q =\frac{1}{p V} \sum_{\boldsymbol{x} \boldsymbol{y}} \overline{ \langle  \epsilon_{\boldsymbol{x}}  s_{\boldsymbol{x}}
     \epsilon_{\boldsymbol{y}} s_{\boldsymbol{y}}\rangle^q}\,,
    \label{eq:def_chiq}
\end{equation}
which scale with $L$ as
\begin{equation}\label{eq:chiq-scaling-1}
  \chi_q  \sim \int^L d^D x ~C_q(r) \sim L^{D-\tau(q)}\,,
 \end{equation}
provided that $D>\tau(q)$. The above relation allows us to  to compute $\tau(q)$ and, from it, $\zeta(q)$. If $D<\tau(q)$ we have that 
\begin{equation}\label{eq:chiq-scaling-2}
\chi_q \sim L^0\,\,\,. 
\end{equation}
Equations~\eqref{eq:chiq-scaling-1} and~\eqref{eq:chiq-scaling-2} provide a very direct test  for multiscaling. Indeed, given that $\tau(1)=1.036(1)$ in three dimensions, one notices that $q\tau(1) > D$ whenever $q\geq 3$. Hence finding (as we shall do below) that $\chi_{q=3}$ scales with a positive power of $L$ gives a direct confirmation of multiscaling behavior.

 \subsubsection{Local susceptibilities}
 
Another strategy is based on computing a different "local susceptibility" $\tilde{\chi}_q$ defined as
\begin{equation}
\tilde{\chi}_q = \frac{1}{p V} \sum_{\boldsymbol{x}} \overline{\chi_{\boldsymbol{x}}^q}\,,
\end{equation}
with
\begin{equation}\label{eq:local-chi}
\chi_{\boldsymbol{x}} = \sum_{\boldsymbol{y}} \langle \epsilon_{\boldsymbol{x}} 
   s_{\boldsymbol{x}} \epsilon_{\boldsymbol{y}}  s_{\boldsymbol{y}} \rangle\,.
\end{equation}
The scaling of this observable, see  \ref{app:scaling}, is such that 
\begin{equation}
   R^{(1)}_q \equiv  \frac{\tilde{\chi}_q}{\tilde{\chi}_1^q}\sim L^{(D-2 +\eta)(q-\zeta(q))/2}\,.
    \label{eq:scalingR1}
\end{equation}
Notice that $\tilde{\chi}_1=\chi_1$.

We have been able to compute numerically the first two derivatives of $\zeta(q)$ by computing ($q\ge 2$)
\begin{equation}
  R^{(2)}_q\equiv   \frac{\tilde{\chi}_q}{\tilde{\chi}_{q-1}} \sim L^{(D-2 +\eta)(\zeta(q-1)-\zeta(q))/2}\
   \label{eq:scalingR2}
\end{equation}
and (for the discrete proxy of the second derivative)
\begin{equation}
    R^{(3)}_q\equiv   \frac{\tilde{\chi}_q \tilde{\chi}_{q-2} }{\tilde{\chi}_{q-1}^2} \sim L^{(D-2 +\eta)(2 \zeta(q-1)-\zeta(q)-\zeta(q-2))/2}\ \,.
    \label{eq:scalingR3}
\end{equation}

\subsection{Details about the computation of the observables}\label{sect:observables-computation}

Since the computation of the correlation function $C_q(r)$ technically differs from the computation of the global susceptibility $\chi_q$ (that, on its side, is also very different from the computation of the local susceptibility  $\tilde{\chi_q}$), we have chosen to discuss these points in separate paragraphs. 

For the computation of $\chi_q$ and $\tilde{\chi}_q$ we shall be employing \emph{real replicas} for every sample. Real replicas are statistically independent system copies that evolve under the same quenched disorder $\{\epsilon_{\boldsymbol{x}}\}$.

Furthermore, our computation of the two susceptibilities uses the enhanced cluster estimator~\cite{sokal:97}
\begin{equation}\label{eq:C-cluster-estimator}
\langle  \epsilon_{\boldsymbol{x}}  s_{\boldsymbol{x}}
     \epsilon_{\boldsymbol{y}} s_{\boldsymbol{y}}\rangle =\langle \gamma_{\boldsymbol{y},\boldsymbol{x}}\rangle\,,
\end{equation}
where $\gamma_{\boldsymbol{y},\boldsymbol{x}}=1$ if, and only if:
\begin{itemize}
    \item $\epsilon_{\boldsymbol{x}}=\epsilon_{\boldsymbol{y}}=1$,
    \item When the lattices is decomposed on Fortuin-Kasteleyn clusters, both sites $\boldsymbol{x}$ and $\boldsymbol{y}$ happen to belong to the same cluster.
\end{itemize}
Otherwise, $\gamma_{\boldsymbol{y},\boldsymbol{x}}=0$.

\subsubsection{The computation of $\chi_q$.\label{subsubsection:chiq}}
Our computation of the global susceptibilities $\chi_q$ uses 12 real replicas. Let  $\gamma_{\boldsymbol{y},\boldsymbol{x}}^{(a)}$ be the enhanced estimator for correlation functions obtained for replica $a$, $(a=1,2,\ldots,12)$. 

Our estimator of $\chi_q$ is obtained in the following way. We start by choosing randomly a starting site $\boldsymbol{x}$ and trace the cluster to which $\boldsymbol{x}$ belongs in all the 12 replicas. Next, we select a set of $q$ distinct replicas $a_1, a_2, \ldots,a_q$, and compute
\begin{equation}
M_{\boldsymbol{x}}^{(q)}=\sum_{\boldsymbol{y}}\, \Big(\prod_{j=1}^q \gamma_{\boldsymbol{y},\boldsymbol{x}}^{(a_j)}\Big)\,.   
\end{equation}
Defining the number of spins as
\begin{equation}
    N_{\text{spins}}=\sum_{\boldsymbol{x}} \epsilon_{\boldsymbol{x}}\,, 
\end{equation}
one easily finds that
\begin{equation}\label{eq:estimator-Fede}
\frac{1}{N_\text{spins}} \sum_{\boldsymbol{x}} \langle M_{\boldsymbol{x}}^{(q)}\rangle=\frac{1}{N_\text{spins}} \sum_{\boldsymbol{x} \boldsymbol{y}} \overline{ \langle  \epsilon_{\boldsymbol{x}}  s_{\boldsymbol{x}}
     \epsilon_{\boldsymbol{y}} s_{\boldsymbol{y}}\rangle^q}\,.
\end{equation}
Of course, in order to improve the statistical accuracy, one can average over different choices of the subset of $q$ distinct replica indices. Note that in practice we choose just one starting site $\boldsymbol{x}$ in Eq.~\eqref{eq:estimator-Fede}. This starting site is picked with uniform probability $1/N_\text{spins}$.

Now, given that
\begin{equation}
    \overline{N_\text{spins}}=p V \,,
\end{equation}
and that the fluctuations of $N_\text{spins}$ do not play any relevant role,\footnote{More quantitatively, trading $1/pV$ with $1/N_{\text{spins}}$ only induces additional corrections to scaling, of size $L^{-a}$ with $a=D-\frac{1}{\nu}\geq D/2$. This estimate can be obtained by combining the following three observations: (i) $\overline{(N_{\text{spins}} -p V)\langle O\rangle}=p(1-p)\frac{\partial{\overline{\langle O\rangle}}}{\partial p}$~\cite{ballesteros:98d},
(ii) $(N_\text{spins}-pV)\sim V^{1/2}$ which justifies the Taylor expansion 
\(\frac{1}{N_{\text{spins}}}=
\frac{1}{pV+(N_{\text{spins}} -p V)}=\frac{1}{pV} \Big(1-\frac{N_{\text{spins}} -p V}{pV}+\ldots\Big)\) and (iii) the Finite-Size Scaling relation $\overline{\langle O \rangle}= L^{x_O/\nu} \mathcal{F}_O(L^{1/\nu}(p-0.8))$  holds when one works precisely at the critical temperature for $p=0.8$ ($x_O$ is the scaling dimension for $O$).} we may approximate
\begin{equation}
    \chi_q\approx \overline{\langle M_{\boldsymbol{x}}^{(q)}\rangle}\,.
\end{equation}
A final note of warning is in order. We cannot employ in the update of the spin configuration  the same Fortuin-Kasteleyn clusters that we employ in the computation of the susceptibilities. Indeed, our twelve real-replicas are to remain statistically independent at all times. Hence we cannot start the single-cluster update from the same site $\boldsymbol{x}$ in all the replicas. This is why we separate our simulations in two different phases:
\begin{enumerate}
    \item During the update phase, we chose independently the starting site $\boldsymbol{x}$ for the single-cluster update of each replica. The spin-configuration is updated after the cluster is traced.
    \item During the measuring phase, instead, the starting point for the cluster tracing is the same in all replicas. However, the spin configurations of the replicas are \emph{not} modified after the $M_{\boldsymbol{x}}^{(q)}$ estimators are computed.
\end{enumerate} 
In both phases, the starting site for cluster tracing is chosen to be occupied (namely, $\epsilon_{\boldsymbol{x}}=1$).

\subsubsection{The computation of $\tilde{\chi}_q$}

Our computation of the local susceptibilities $\tilde{\chi}_q$ employs 16 real replicas. The separation between update phase and measuring phase that we have discussed above also applies to this case. 

During the measuring phase, we compute for each replica
\begin{equation}
N_{\boldsymbol{x}}^{(a)}= \sum_{\boldsymbol{y}} \gamma_{\boldsymbol{y},\boldsymbol{x}}^{(a)}\,.
\end{equation}
We proceed by selecting a subset of $q$ distinct replica indices  $a_1, a_2, \ldots,a_q$, and compute the estimator
\begin{equation}
Y_{\boldsymbol{x},q}=\prod_{j=1}^q N_{\boldsymbol{x}}^{(a_j)}\,,
\end{equation}
which has the thermal expectation value
\begin{equation}
    \langle Y_{\boldsymbol{x},q}\rangle=\chi_{\boldsymbol{x}}^q\,,
\end{equation}
where the local susceptibility $\chi_{\boldsymbol{x}}$ was defined in Eq.~\eqref{eq:local-chi}. Of course, one may average over the different choices of the subset of $q$ distinct replicas in order to increase the statistics.

As in the previous paragraph, we pick just one starting point $\boldsymbol{x}$ during the measuring phase, with uniform probability $1/N_{\text{spins}}$. Hence
\begin{equation}
\tilde{\chi}_q\approx \overline{\langle Y_{\boldsymbol{x},q}\rangle}\, .
\end{equation}
The above equation is not an equality because of the fluctuations on the number of spins $N_{\text{spins}}$.

\subsubsection{The computation of $C_q(r)$}
As opposed to the strategy we have followed in the computation of the susceptibilities based in the simulation of a high number of replicas (in the same realization of the disorder), in the computation of the correlation function $C_q$ we have chosen to simulate only one replica. This greatly simplifies the simulation but induces strong bias in the statistical estimator of $C_q$. In particular, it is possible to show~\cite{ballesteros:98} that this bias is proportional to $1/N_\mathrm{m}$ ($N_\mathrm{m}$ being the number of measurements on a given sample), and the problems arise  when the magnitude of this bias is  similar to the statistical error induced by the disorder (proportional to $1/\sqrt{N_\mathrm{S}}$ ($N_\mathrm{S}$ is the number of disorder realizations, samples).

To cure this bias, we have followed the strategy introduced in Ref.~\cite{ballesteros:98}. Let us work on a given sample (i.e. fixed disorder) and at equilibrium. Our strategy consists, firstly, in computing the total average of a given observable $O$, denoted as $O^{(1)}$. Then we take two halves of the Monte Carlo simulation, and compute the average value of $O$ in these two halves and compute their mean, denoted as $O^{(2)}$. Finally, obtain the average of $O$ on each quarter of the Monte Carlo history and compute the mean of them, getting $O^{(3)}$. The final, unbiased, quadratic estimator is~\cite{ballesteros:98}
\begin{equation}
O^Q=\frac{8}{3} \, O^{(1)}-2 \, O^{(2)}+\frac{1}{3} \, O^{(3)}\,.
\end{equation}
Finally, in Table \ref{tab:stat} we report the number of samples used in the different runs.

\begin{table}[tb]
\caption{Details of our numerical simulations. We report the number of samples (disorder realizations), $N_\mathrm{S}$, used in the different runs ($\chi_q$ susceptibilities, $\tilde{\chi}_q$ local susceptibilities and correlation functions). We also report the number of sweeps (as defined in the text) used for thermalizing the different systems, $N_\mathrm{sweeps}$. In the  $\chi_q$ susceptibility runs we have simulated 12 replicas per sample while for computing  the local susceptibilities we have simulated 16 replicas per sample. In the correlation (Corr.) runs we have simulated only one replica per sample.}
\begin{indented}
\item[]\begin{tabular}{@{}c  c  c  c  c}
\br
$L$ & $N_\mathrm{S}$ ($\chi_q$) &  $N_\mathrm{S}$ ($\tilde{\chi}_q$) & $N_\mathrm{S}$ (Corr.) & $N_\mathrm{sweeps}$\\
\mr
8 &         & 110592 &  &16\\
12 & 98304 &  108902 &  &32\\
16 & 98304 & 109666 &  & 32\\
24 & 98304 & 109428 &  & 64\\
32 & 81920 & 796367 & 5000 & 64\\
48 & 81920 & 109528 & 5030 & 128\\
64 & 81920 & 84774 & 9800  & 128\\  
96 & 81920 & 84034 &  & 512\\
128 & 26984 & 60015 &  & 512\\
\br
\end{tabular}
\label{tab:stat}
\end{indented}
\end{table}

\section{Numerical Results}\label{sect:results}

In this section we  present our numerical results regarding multiscaling by analyzing the behavior of the correlation functions and the global and local susceptibilities.

\subsection{Correlation functions}

\begin{figure}[h]
 \includegraphics[width=\columnwidth]{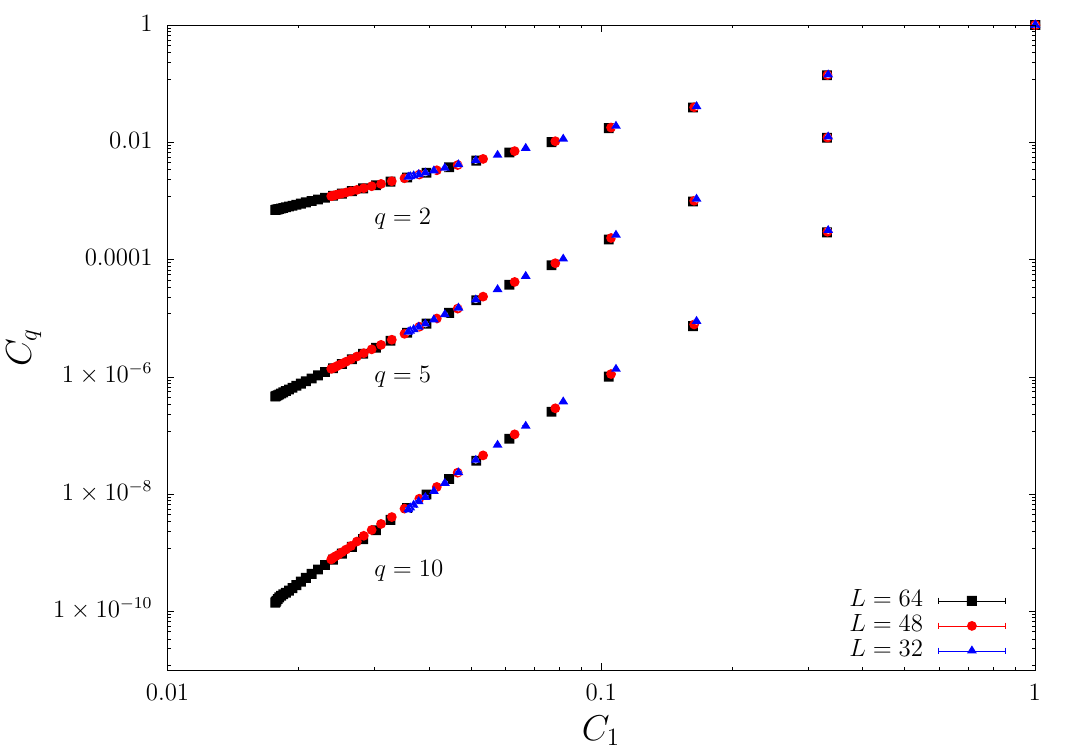}
  \caption{ $C_q(r)$ versus $C_1(r)$ for  $q=2$, 5 and 10, and  for $L=32$ (blue triangles), $L=48$ (red circles) and $L=64$ (black squares). The size of the error bars is smaller than the ones of the symbols.}
  \label{fig:scaling}
\end{figure}

To quantify the $\zeta(q)$ exponent we have analyzed the dependence of $C_q$ on $C_1$. We show in Fig. \ref{fig:scaling} the behavior of $C_q$ as a function of $C_1$ for three different lattice sizes and three values of $q$. The scaling of the data (for $L=32$, 64 and 128) is very good and the power law dependence of $C_q$ on $C_1$ is very clear and accurate. 

\begin{figure}[h]
 \includegraphics[width=\columnwidth]{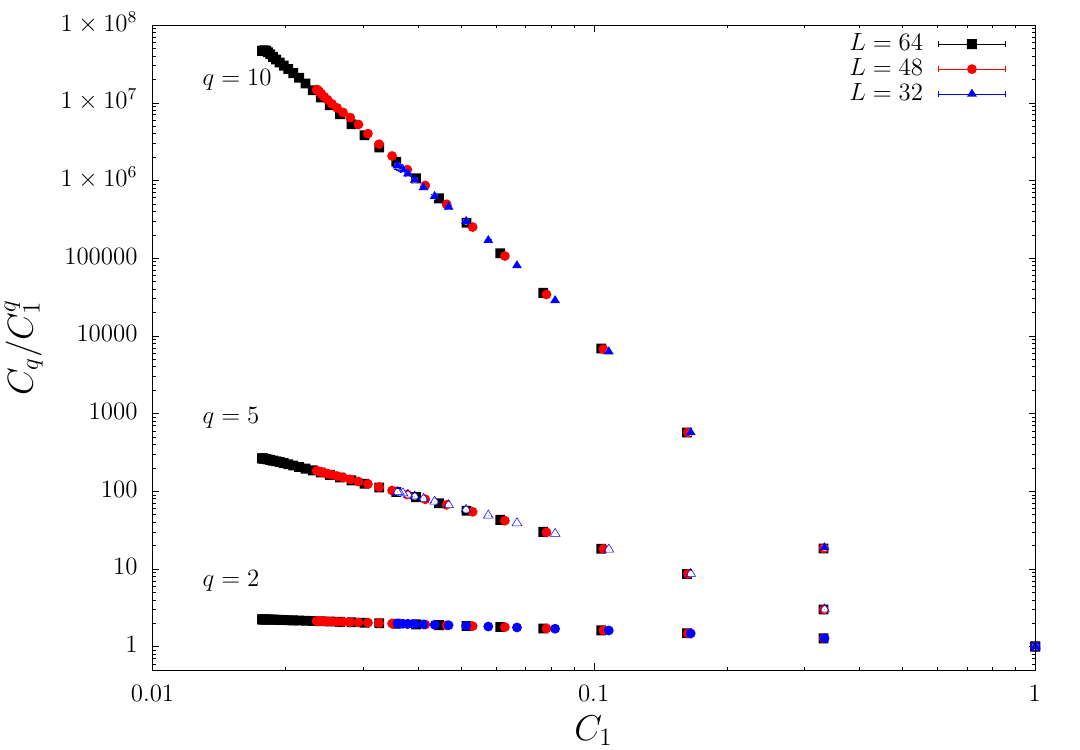}
  \caption{ $C_q(r)/C_1^q(r)$ versus $C_1(r)$ for  $q=2$, 5 and 10, and  for $L=32$ (blue triangles), $L=48$ (red circles)  and $L=64$ (black squares). Notice the potential growth in all the cases which is a clear signal of the multiscaling behavior. All the curves take the value 1 as $C_1=1$ (the rightmost point of the plot). The error bars of the ratio have been computed using the jackknife method. The size of the error bars is smaller than the ones of the symbols.}
  \label{fig:multiscaling}
\end{figure}

In Fig. \ref{fig:multiscaling} we plot $C_q/C_1^q$ as a function of $C_1$ in order to asses the multiscaling behavior. In all the simulated cases ($q\le 10$) we have found a diverging behavior as $C_1 \to 0$ which marks the onset of multiscaling behavior in this model.  Notice again the good scaling of the data from different lattice sizes ($L=32$, 48 and 64).

We can estimate the  $\zeta(q)$ exponent by fitting the correlation function data to Eq. (\ref{eq:beta}). We fit $C_q(r)$ computed on a given lattice size against $C_1(r)$ computed on the same lattice size. We have computed the error bars of the $\zeta(q)$ using the jackknife method over the number of samples, following the recipe of Ref.~\cite{Yllanes:11}: we perform independent fits on all the jackknife blocks using the diagonal covariance matrix, and then,  the error bars in the parameters of the fit can be computed from the fluctuations among all the jackknife blocks.

We report our results in Table \ref{tab:power}. The exponent $\zeta(q)$  is not proportional to $q$. This  clearly shows that the three-dimensional diluted Ising model presents multiscaling behavior. Despite the fact that we are using a quasi-perfect action, we can observe a small dependence of $\zeta(q)$ on the lattice size.

\begin{table}[tb]
\caption{ Exponent $\zeta(q)$ computed using the $L=32$, $L=48$ and $L=64$ correlation functions. This exponent  has been computed using the jackknife method to tackle the highly correlation of the data (different $r$ in the correlations $C_q(r)$). We also report the $C_1$-interval used in the fits,  $(C_1^{\mathrm{m}},C_1^{\mathrm{M}})$.}
\label{tab:power}
\centering
\begin{indented}
\item[]
\begin{tabular}{c cc cc cc}
\br
 &\multicolumn{2}{c}{$L=64$} & \multicolumn{2}{c}{$L=48$} & \multicolumn{2}{c}{$L=32$}\\
\br
$q$ & $(C_1^{\mathrm{m}},C_1^{\mathrm{M}}) $ & $\zeta(q)$   & $(C_1^{\mathrm{m}},C_1^{\mathrm{M}}) $ & $\zeta(q)$  &
$(C_1^{\mathrm{m}},C_1^{\mathrm{M}}) $ & $\zeta(q)$ \\
\br
2 & (0,0.3) &1.8109(3) &   (0,0.3) & 1.8094(7) &(0,0.3) &1.808(1) \\
3 & (0,0.17) & 2.457(1)& (0,0.3) & 2.4525(19) &(0,0.3) &2.447(3) \\
4 & (0,0.1) & 3.026(4) &  (0,0.16) &2.997(6) &(0,0.3) &2.993(8)\\
5 & (0,0.1) & 3.486(7) &  (0,0.1) & 3.474(13) &(0,0.3) &3.432(12)\\
6 & (0,0.1) & 3.849(9) &  (0,0.1) & 3.836(18) &(0,0.3) &3.803(18)\\
7 & (0,0.1) & 4.192(13) &  (0,0.1) & 4.17(2) &(0,0.3) &4.23(3)\\
8 & (0,0.1) & 4.483(17) &  (0,0.1) & 4.45(3) &(0,0.3) &4.54(5)\\
9 & (0,0.1) & 4.75(2) &  (0,0.1) & 4.72(4) &(0,0.3) &4.83(6)\\
10 & (0,0.1) & 5.00(3) &  (0,0.1) &4.95(5) &(0,0.3) &5.10(7)\\
\br
\end{tabular}
\end{indented}
\end{table}

\subsection{Global Susceptibilities}

\begin{figure}[h]
 \includegraphics[width=\columnwidth]{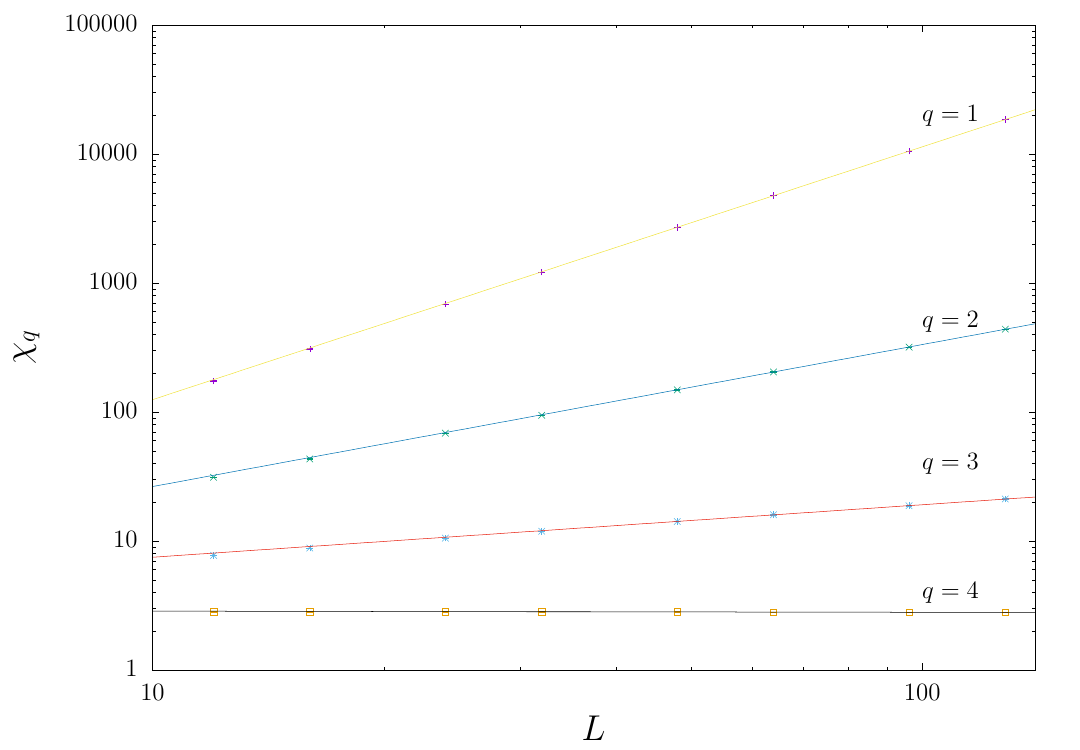}
  \caption{ $\chi_q$ versus $L$ for  $q=1$, 2, 3 and 4 in a double logarithm scale. The straight lines are fits to Eq. (\ref{eq:chiq-scaling-1}) using $L\ge 48$ to avoid the small scaling corrections. For $q \ge 4$, $\chi_q$ does not diverge, so $\tau(q) \ge D=3$. In order to compute $\zeta(q)$ we have rescaled $\tau(q)$ using the numerical value for $1+\eta=1.036(1)$ (see Eq. (\ref{eq:tautheta})): so $\zeta(q)=\tau(q)/(1+\eta)=\tau(q)/1.036(1)$.}
  \label{fig:sus_scaling}
\end{figure}

Figure \ref{fig:sus_scaling} shows the behavior of the $\chi_q$ susceptibilities as a function of the lattice size.

By fitting the susceptibilities presented in Fig. \ref{fig:sus_scaling} we have estimated the $\tau(q)$ exponents using Eq. (\ref{eq:chiq-scaling-1}). The numerical estimates for the $\tau(q)$ exponents (and consequently those for the $\zeta(q)$ exponents) can be found in Table \ref{tab:expsus}.

\begin{table}
\caption{$\tau(q)$ and $\zeta(q)$ exponents from the finite size scaling of the susceptibilities $\chi_q$. For $\chi_4$ we do not detect divergence with $L$, so for $q\ge 4$,  $\tau(q)\ge 3$ and   $\theta(q) \ge 2.89(1)$. We have computed, as a test,  the $1+\eta$ exponent in the $q=1$ row, and it compares very well with the most accurate available estimate $1+\eta=1.036(1)$~\cite{hasenbusch:07}.}
\begin{indented}
\item[] \begin{tabular}{c c c } 
\br
$q$ & $\tau(q)$ & $\zeta(q)$\\
\mr
1 & 1.039(2) &  1 (by def.)\\
2 & 1.893(3) & 1.833(5) \\
3 & 2.586(4) &  2.503(6) \\
4 &  $\ge 3$ &  $\ge 2.896(3)$\\
\br
\end{tabular}
\label{tab:expsus}
\end{indented}
\end{table}

\subsection{Local susceptibilities}

\begin{figure}[h]
 \includegraphics[width=\columnwidth]{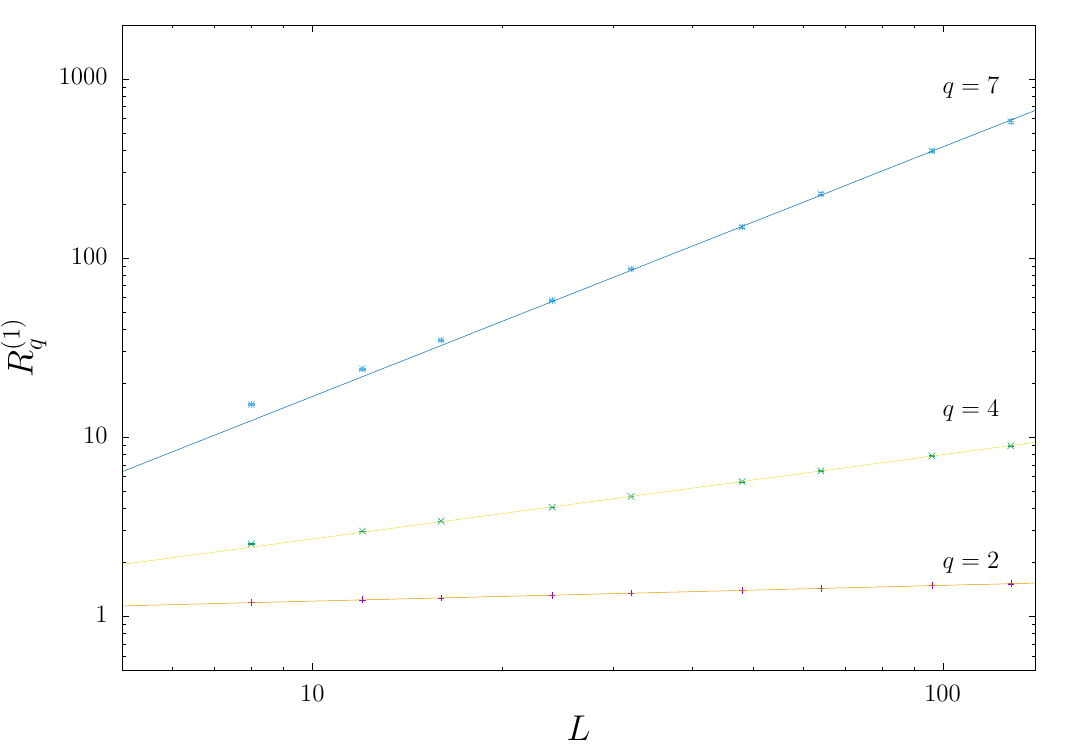}
  \caption{ $R^{(1)}_q$ versus $L$ for  $q=2$, 4, and 7 in a double logarithm scale. The straight lines are fits to Eq. (\ref{eq:scalingR1}) considering scaling corrections. The error bars of all three ratios $R^{(i)}_q$ have been computed using the bootstrap method.}
  \label{fig:sus_scaling_loc}
\end{figure}

In Fig. \ref{fig:sus_scaling_loc} we show the behavior of $R^{(1)}_q$ versus $L$ for three different values of $q$,  showing the power law divergence implied by Eq. (\ref{eq:scalingR1}).

\begin{figure}[h]
 \includegraphics[width=\columnwidth]{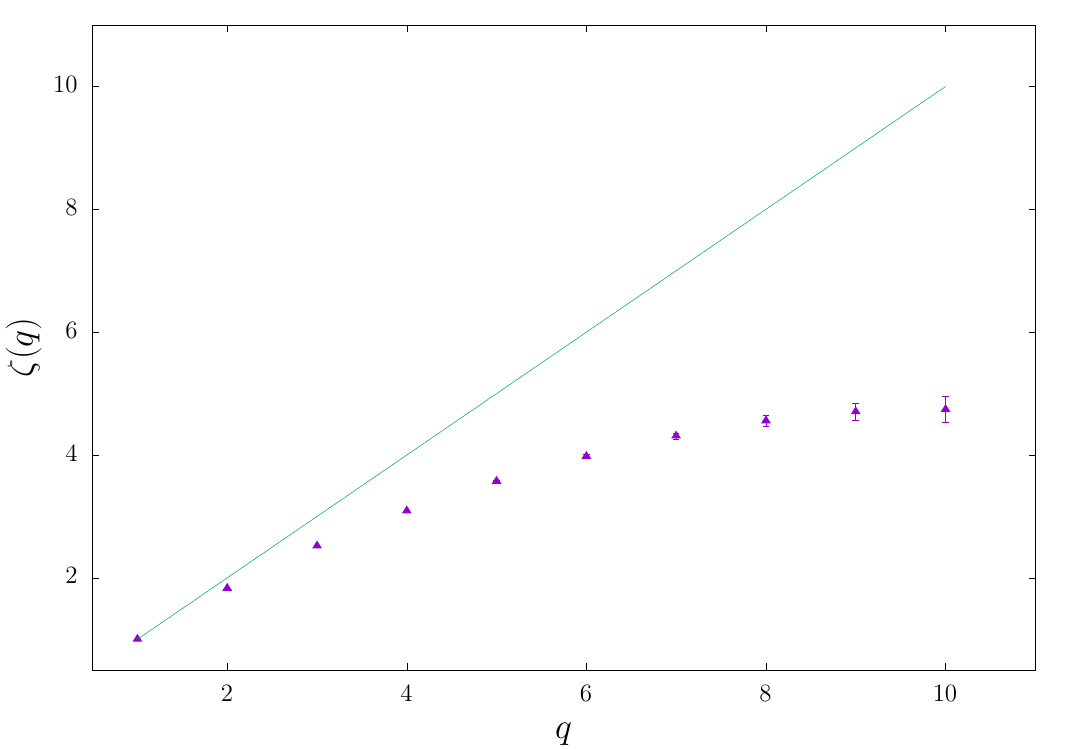}
  \caption{The exponent $\zeta(q)$ versus $q$ (see results of Table \ref{tab:expsusloc}). By definition $\zeta(1)=1$. We have also plotted the line $\zeta=q$ to show the appearance of the multi-fractal behavior in the model ($\zeta(q) \neq q$).}
  \label{fig:zetaq}
\end{figure}

In Table \ref{tab:expsusloc} we present our final values for the $\zeta$-exponents and the (discrete) first and second derivatives obtained analyzing the observables $R^{(1)}_q$, $R^{(2)}_q$ and $R^{(3)}_q$, computing their error bars with the bootstrap method, and using 
only the lattice sizes with $L\ge 48$ (as for the global susceptibility analysis) to avoid the small scaling corrections~\footnote{An analysis using all the sizes and assuming the leading scaling correction term gives essentially the same exponents.}. We plot our results for $\zeta(q)$ in Fig. \ref{fig:zetaq} where the strong departure of the linear regime is clear, providing strong indications of multi-fractal behavior. 

Furthermore, the results reported in Table \ref{tab:expsusloc} show that the function $\zeta(q)$ is a non-decreasing function for the values of $q$ simulated. In addition, it is a concave function of $q$, in agreement with the results from Ludwig~\cite{ludwig:89} (see also  \ref{app:convex}).

\begin{table}[tb]
\caption{$\zeta(q)$ exponents and the  (discrete) first and second derivatives from the finite size scaling of the ratios $R^{(1)}_q$, $R^{(2}_q$ and $R^{(3)}_q$ of the local  susceptibilities $\tilde{\chi}_q$. Notice $\zeta(1)=1$.}
\begin{indented}
\item[]\begin{tabular}{c c c c} 
\br
$q$ & $\zeta(q)$  & $\zeta(q)-\zeta(q-1)$  &$\zeta(q)+\zeta(q-2)-2 \zeta(q-1)$ \\
\mr
2 & 1.830(2)&   0.830(2)&                  \\
3 & 2.517(5)&    0.688(3) &   -0.143(2)     \\
4 & 3.09(1)&   0.573(6) &   -0.115(3)      \\
5 & 3.57(2)&   0.481(9)  &   -0.092(4)     \\        
6 & 3.98(3)&     0.40(1) &    -0.077(6)     \\   
7 &  4.31(5)&    0.33(2)  &    -0.07(1)     \\   
8 &  4.56(8)&    0.25(4)  &    -0.08(2)     \\   
9 &  4.7(1)&   0.16(6)  &    -0.09(2)      \\ 
10 & 4.7(2) &    0.05(9)  &     -0.10(3)     \\   
\br
\end{tabular}
\label{tab:expsusloc}
\end{indented}
\end{table}

\section{Discussion and Conclusions}\label{sect:conclusions}

Along the years, many disordered ferromagnetic systems  in two dimensions have been shown to exhibit multiscaling behavior (see e.g. Ref.~\cite{ludwig:89}), with the surprising exception of the DIM. This is however
a marginal behavior. The DIM is the $Q=2$ instance of the  diluted Potts model with $Q$ states (which does display multiscaling for $Q>2$). Furthermore, it suffices to consider long range interactions to find multiscaling in the two-dimensional DIM (see~\cite{chippari:23} and references therein). The behavior of the DIM (with short range interactions) in $D=2$ is  marginal in a different way, as well. Although there is no multiscaling in $D=2$, in a pioneering work, Davis and Cardy~\cite{davis:00}  found  multifractality in the DIM in space dimensions $D=2+\epsilon$. Here, we have extended Davis and Cardy's~\cite{davis:00} result to $D=3$ through extensive numerical simulations for the site-diluted Ising model, that is studied under equilibrium conditions. 

We have computed the exponents $\tau(q)$ for integer values $q=1,2,\ldots,10$. The $q$-th power of the correlation function, Eq.~\eqref{eq:defCK}, decays at long distances as $1/r^{\tau(q)}$~\eqref{eq:defTheta}. We have found a very clear numerical evidence for $\tau(q)<q\tau(1)$ for $q>1$, which is the hallmark of multifractality.

Let us emphasize that multiscaling in no way contradicts the general Renormalization Group expectation of a universal large-$L$ limit for relative cumulants of the order-parameter~\cite{aharony:96}. Specifically, for the very same model studied here, it was shown long ago that the relative variance of the squared order parameter, \begin{equation}
g_2=\frac{\overline{\langle \mathcal{M}^2\rangle^2}\,-\,\overline{\langle \mathcal{M}^2\rangle}^2}{\overline{\langle \mathcal{M}^2\rangle}^2}\, \quad \text{ with }\quad
\mathcal{M}=\sum_{\boldsymbol{x}}\, \epsilon_{\boldsymbol{x}}  s_{\boldsymbol{x}}\,,
\end{equation}
reaches its expected universal limit~\cite{ballesteros:98b} (this behavior is sometimes named no self-averaging~\cite{aharony:96}). One could think that having a finite large-$L$ limit for $g_2$ does somehow contradict multiscaling. The correct conclusion is indeed different,  as we now explain starting from the results discussed in Appendix~\ref{app:scaling}. Consider the second moment, $q=2$, which is one relevant in the computation of $g_2$. The key point regards the spatial sites one integrates over:
\begin{itemize}
    \item If one integrates $\overline{\epsilon_{\boldsymbol{x}} \epsilon_{\boldsymbol{y}}\langle S_{\boldsymbol{x}} S_{\boldsymbol{y}}\rangle^2}$ over $\boldsymbol{y}$ the global susceptibility $\chi_{q=2}$ is obtained and the full multifractal signal is recovered.
\item If one integrates instead 
\(\overline{\epsilon_{\boldsymbol{x}} \epsilon_{\boldsymbol{y}_1}\epsilon_{\boldsymbol{y}_2}\langle S_{\boldsymbol{x}} S_{\boldsymbol{y}_1}\rangle\langle S_{\boldsymbol{x}} S_{\boldsymbol{y}_2}\rangle}\)
over $\boldsymbol{y}_1$ and $\boldsymbol{y}_2$ the resulting quantity is the local susceptibility $\tilde{\chi}_{q=2}$ which has only half the multifractal signal.\footnote{$\tilde{\chi}_q$  more than compensate this disadvantage by scaling as a positive power of $L$ for all $q\geq 1$.}
\item Yet, the needed quantity in the computation of $g_2$ is $\overline{\langle\mathcal{M}^2\rangle^2}$. Hence, we are to integrate \(\overline{\epsilon_{\boldsymbol{x}_1} \epsilon_{\boldsymbol{x}_2}\epsilon_{\boldsymbol{y}_1}\epsilon_{\boldsymbol{y}_2}\langle S_{\boldsymbol{x}_1} S_{\boldsymbol{y}_1}\rangle\langle S_{\boldsymbol{x}_1} S_{\boldsymbol{y}_2}\rangle}\) over $\boldsymbol{x}_1$, $\boldsymbol{x}_2$, $\boldsymbol{y}_1$ and $\boldsymbol{y}_2$. But the integrals over $\boldsymbol{x}_2$, $\boldsymbol{y}_1$ and $\boldsymbol{y}_2$ already suffice to completely suppress multifractal scaling. In fact, neglecting scaling corrections, $\overline{\langle\mathcal{M}^2\rangle^2}$ scales with $L$ in the same way that $\overline{\langle\mathcal{M}^2\rangle}^2$ does.
\end{itemize}
In summary, the extreme self-averaging violations evinced by the multiscaling analysis can only be identified when studying correlations at the \emph{local} level. Carrying out too many spatial integrations erases the multifractal signal. In fact, to unearth multiscaling we have pursued here the same local approach that has been used in Ref.~\cite{janus:23c} in the context of the out-equilibrium dynamics of spin-glasses~\footnote{However, it is worth reminding that the spin glass criticality is much less understood for the lack of a proper and easy-to-use renormalization group theory \cite{angelini2013ensemble,angelini2017real,lubensky2023renormalization}.}.

\section*{Acknowledgments}

We are thankful to Marco Picco for useful discussions. We also thank Maria Chiara Angelini for discussions about a related problem in the context of site percolation.

Our simulations have been carried out at the the \textit{Instituto de Computación Científica Avanzada de Extremadura} (ICCAEx), at Badajoz,  We would like to thank its staff. 
This work was partially supported by Ministerio de Ciencia, Innovaci\'on y Universidades (Spain), Agencia Estatal de Investigaci\'on (AEI, Spain, 10.13039/501100011033), and European Regional Development Fund (ERDF, A way of making Europe) through Grants PID2020-112936GB-I00 and PID2022-136374NB-C21, by the Junta de Extremadura (Spain) and Fondo Europeo de Desarrollo Regional (FEDER, EU) through Grant No.\ IB20079.
This research has also been supported by the European Research Council under the European Unions Horizon 2020 research and innovation program (Grant No. 694925—Lotglassy, G. Parisi) and by ICSC—Centro Nazionale di Ricerca in High Performance Computing, Big Data, and Quantum Computing funded by European Union—NextGenerationEU.
EM acknowledges funding from the PRIN funding scheme
(2022LMHTET - Complexity, disorder and fluctuations: spin glass physics and beyond) and from the FIS (Fondo Italiano per la Scienza) funding scheme (FIS783 - SMaC - Statistical Mechanics and Complexity: theory meets experiments in spin glasses and neural networks) from Italian MUR (Ministery of University and Research).

\appendix
\renewcommand{\thesection}{\Alph{section}}
\section{Some theoretical results}

\subsection{Scaling dimensions} \label{app:scaling}

In the replicated theory the $C_q$ correlation functions can be written as
\begin{equation}
\langle (\phi^{a_1}(\boldsymbol{x})\cdots \phi^{a_q}(\boldsymbol{x})) ~ (\phi^{a_1}(\boldsymbol{y})\cdots \phi^{a_q}(\boldsymbol{y})) \rangle \,,
\end{equation}
with $1 \le a_i<a_j \le n$ for $i<j$, where $\phi^{a}(\boldsymbol{x})$ are the replicated fields with scaling dimension $[(D-2+\eta)/2]$ and $n$ is the number of replicas ($n\to 0$). 

As discussed in Ref.~\cite{ludwig:89}, the composite operator $(\phi^{a_1}(\boldsymbol{x})\cdots \phi^{a_q}(\boldsymbol{x}))$ transforms following a reducible representation of the symmetric group ($S_n$) and, therefore, it is not a scaling operator at the random critical point. However, it can be expressed as a linear combination of scaling fields $\Phi_q^{{\mu},\alpha}(\boldsymbol{x})$, and each of them  transforms following the $\mu$-
irreducible representation (irrep) of $S_n$ ($\alpha$ labels the multiplicity of the $\mu$-irrep)~\cite{ludwig:89}, with scaling dimension $X_q^{{\mu},\alpha}$. Its asymptotic behavior is
\begin{equation}
\langle \Phi_q^{{\mu},\alpha}(\boldsymbol{x})  \Phi_q^{{\mu},\alpha}(0) \rangle \sim 
\frac{1}{|\boldsymbol{x}|^{2 X_q^{\mu,\alpha}}}\,.
\end{equation}
Therefore, the behavior of $C_q$ will be a sum (over the irreducible representations) of decays with powers $1/|\boldsymbol{x}|^{2 X_q^{\mu,\alpha}}$, 
and the smallest scaling dimension $X_q^{{\mu},\alpha}$ will determine its large scale behavior. 

It is possible to show that for 
the spin operator (which is our case), all the different representations become degenerate as opposed to the energy operator~\cite{ludwig:89, davis:00}. Therefore, 
$[(D-2+\eta)/2] \zeta(q)$ is the scaling dimension of the {\em full} composite operator 
$(\phi^{a_1}(\boldsymbol{x})\cdots \phi^{a_q}(\boldsymbol{x}))$, which defines $\zeta(q)$ in such a way that $\zeta(1)=1$.

Hence, the scaling behavior of $C_q$ will be twice that of the composite operator $ (\phi^1(\boldsymbol{x})\cdots \phi^q(\boldsymbol{x}))$, namely $[(D-2+\eta)/2] \zeta(q) \times 2$.

In the case of the local susceptibilities $\tilde{\chi}_q$ we need to compute
\begin{equation}
\frac{1}{V} \sum_{\boldsymbol{x}} \sum_{\boldsymbol{y}_1}\cdots \sum_{\boldsymbol{y}_q} \overline{\langle S_{\boldsymbol{x}} S_{\boldsymbol{y}_1}\rangle \cdots \langle S_{\boldsymbol{x}} 
S_{\boldsymbol{y}_q}\rangle } \,,
\label{eq:inter}
\end{equation}
that in terms of the replicated theory, the associated correlation function  can be written as
\begin{equation}
\langle (\phi^{a_1}(\boldsymbol{x})\cdots \phi^{a_q}(\boldsymbol{x})) ~\phi^{a_1}(\boldsymbol{y}_1) \cdots \phi^{a_q}(\boldsymbol{y}_q) \rangle\,,
\end{equation}
and the scaling dimension of the only one composite operator   $(\phi^{a_1}(\boldsymbol{x})\cdots \phi^{a_q}(\boldsymbol{x}))$, is again  $[(D-2+\eta)/2] \zeta(q)$
and each of the $q$-factors $\phi^a(\boldsymbol{y}_a)$ have $(D-2+\eta)/2$ as their scaling dimension. 

Notice that a  field $\phi$ with  $\mathrm{dim}(\phi)$ as the scaling dimension transforms following the rule  $\phi(b \boldsymbol{x}) =b^{-\mathrm{dim}(\phi)} \phi(\boldsymbol{x})$.

Equation (\ref{eq:inter}) can be written in the continuum as
\begin{eqnarray*}
    \tilde{\chi}_q &\sim \frac{1}{L^D}\int d^Dx~\int d^Dy_1 \cdots \int d^Dy_q\ 
                    \langle (\phi^{a_1}(\boldsymbol{x})\cdots \phi^{a_q}(\boldsymbol{x})) ~\phi^{a_1}(\boldsymbol{y}_1) \cdots \phi^{a_q}(\boldsymbol{y}_q) \rangle\,,
\end{eqnarray*}
where the $q+1$ integrals run in the volume $[a,L]^D$ ($a$ is the lattice spacing). After the change of variables $\boldsymbol{\tilde{x}}=\boldsymbol{x}/L$ and
$\boldsymbol{\tilde{y}}_i=\boldsymbol{y}_i/L$ ($i=1,\dots, q$) we can write
\begin{eqnarray*}
    \tilde{\chi}_q &\sim \frac{1}{L^D} L^{(q+1) D} \frac{1}{L^{\left((D-2+\eta)/2\right) \zeta(q)}} \frac{1}{L^{\left((D-2+\eta)/2\right) q}}\\
    &\times\int d^D \tilde{x}~\int d^D \tilde{y}_1 \cdots \int d^D \tilde{y}_q\ 
    (\phi^{a_1}(\boldsymbol{\tilde{x}})\cdots \phi^{a_q}(\boldsymbol{\tilde{x}})) ~\phi^{a_1}(\boldsymbol{\tilde{y}}_1) \cdots \phi^{a_q}(\boldsymbol{\tilde{y}}_q) \rangle\\
    &\sim L^{(q D - [(D-2+\eta)/2] (\zeta(q)+q)} \,,
\end{eqnarray*}
since the $q+1$ integrals in the tilde variables provided a constant in the large $L$-limit: their integration limits are constrained to the volume  $[a/L,1]^D$ which is $[0,1]^D$ for large $L$.

Taking into account this result, it is easy to get the behaviors given in  Eqs. (\ref{eq:scalingR1}), (\ref{eq:scalingR2}) and  (\ref{eq:scalingR3}).

\subsection{Concavity of the scaling exponents} \label{app:convex}

Ludwig showed that $X_q^{\mu,\alpha}$ is a concave function\footnote{To fix the notation, a concave function in an interval $[a,b]$ satisfies
$f(\lambda x_1 + (1-\lambda ) x_2) \ge \lambda f(x_1) + (1-\lambda) f(x_2)$ for $\lambda\in (0,1)$ and for any two points $x_1$ and  $x_2$ in the interval $[a,b]$.} of $q$~\cite{ludwig:89}. The argument is simple, so we will reproduce here for $\zeta(q)$. Notice that if, for a given $r$, the correlation function for a given sample can take only values in the compact interval $[0,1]$, then its probability density function (over the samples) is determined by its ($q$-th) positive moments, that we have denoted as $C_q$. Moreover these moments are smooth functions of $q$, therefore, the $\zeta(q)$ exponents can also be defined for real $q$. In addition, $C_q^{1/q}$ is a non-decreasing  function of $q$ which implies that $\zeta(q)$ is a  concave functions since  $\zeta(q_1)/q_1 \le \zeta(q_2)/q_2$ as $0<q_2<q_1$.

\subsection{Davis-Cardy's result} \label{app:daviscardy}
In Ref.~\cite{davis:00} Davis and Cardy computed the scaling exponents $X_q^{\mu,\alpha}$ for the DIM in $2+\epsilon$ dimensions. In the notation used in this paper, their result can be written as
\begin{equation}
\label{eq:DC}
q-\zeta(q)=  \left( \frac{q (q-1)}{2}\right) y + O(y^2)\,,
\end{equation}
where $y$ is the RG eigenvalue of the disorder strength which is, at this order, $y=\alpha=O(\epsilon)$ ($\alpha$ is the specific heat exponent of the pure model and $\epsilon=D-2$). This results clearly shows that the DIM in $2+\epsilon$ is expected to undergo a multiscaling behavior. However, in order to have an accurate analytical estimate of the difference $q-\zeta(q)$ for the three-dimensional model one would need to extend this computation to higher orders of $y$~\cite{cardy:96}.

\addcontentsline{toc}{section}{References}
\bibliographystyle{iopart-num}

\providecommand{\newblock}{}

\end{document}